\documentclass[aps]{revtex4}
\usepackage{graphicx}
\newcommand{\kbar}{\mathchar'26\mkern-9muk}
\newcommand{\beq}{\begin{equation}}
\newcommand{\enq}{\end{equation}}
\newcommand{\bea}{\begin{eqnarray}}
\newcommand{\ena}{\end{eqnarray}}
\begin{document}
\title{Directed transport and 
Floquet analysis for a periodically kicked wavepacket at 
a quantum resonance}
\author{Emil Lundh}
\affiliation{Department of Physics, Royal Institute of Technology,
AlbaNova, SE-106 91 Stockholm, Sweden}
\affiliation{Center of Mathematics for Applications, 
P.O. Box 1053 Blindern, NO-0316 Oslo, Norway}
\affiliation{Department of Physics, Ume{\aa} University, 
SE-90187 Ume{\aa}, Sweden\footnote{Present address.}}
\begin{abstract}
The dynamics of a kicked quantum mechanical wavepacket at 
a quantum resonance is 
studied in the framework of Floquet analysis. 
It is seen how a directed current can be created 
out of a homogeneous initial state at certain resonances 
in an asymmetric potential. 
The almost periodic parameter dependence of the current 
is found to be 
connected with level crossings in the Floquet spectrum.
\end{abstract}
\maketitle
\section{Introduction}
The kicked rotor is, due to its relative simplicity, one of the 
most studied and best understood models of chaotic mechanics. 
In essence, it models a particle that is exposed to a sinusoidal 
potential during periodically repeated, delta-function shaped kicks. 
The quantum mechanical version of the kicked rotor has attracted 
special attention lately, due to the realization of the model in 
optical lattices \cite{raizen,duffy,sadgrove}. It turns out that 
there exist two peculiar quantum mechanical effects that distinguish 
the quantum kicked rotor from its classical counterpart. These two, 
mutually exclusive, effects are Anderson localization in momentum 
space, which occurs for generic (almost all) values of the kicking period 
\cite{fishman}, 
and quantum resonances, which occur when the kicking period 
matches a resonance criterion 
\cite{duffy,sadgrove,izrailev,izrailevreview,daley,wimberger,dd}. 
The latter phenomenon, quantum resonances, is the subject of this 
paper.

Classically, the energy growth of the kicked rotor is diffusive, 
i.~e.\ linear. In the quantum case and off resonance, the energy 
growth will also be diffusive but eventually saturate due to 
Anderson localization in momentum space. In contrast, a quantum 
resonance will result in an energy growth that is quadratic in 
time. Such resonances 
occur when the period of the potential kicks on the particle 
matches a rational value times the so-called Talbot time of the 
system. In the context of optical lattices, the corresponding 
frequency is identical to the recoil frequency \cite{lw1}.
It is common to state the resonance criterion in terms of an 
effective Planck constant which 
is defined as a combination of the physical parameters.

We have recently found that a slight generalization of the 
kicked rotor to an asymmetric, sawtooth-shaped potential will, 
combined with quantum resonances, result in a ratchet effect of 
sorts -- a directed current which increases linearly with 
time \cite{lw1}. In general, a classical or quantum particle 
in a flashing periodic potential will not pick up a finite 
velocity even if the potential is asymmetric, i.~e.,
there is no ratchet effect. However, 
quantum resonances may change the situation and allow for a directed 
current. The effect hinges on the fact that a specific momentum 
eigenstate is chosen as the initial state, since a proper averaging 
over all of phase space would necessarily cancel out any directed 
current. Nevertheless, it was seen in Ref.\ \cite{lw1} that 
the proposed setup could give a constant acceleration to an 
initially zero-momentum plane wave.

The spectral properties of the quantum kicked rotor were first 
studied by Izrailev and Shepelyanskii \cite{izrailev, izrailevreview}. 
It was found that the quasienergy spectrum is in general discrete, 
but in the case of a quantum resonance it forms a band structure 
with a continuum of quasienergies. 
This was found to explain the quadratic energy growth. 
In the present paper, the same type of analysis is exploited in order to 
understand how a directed current is created 
out of a motionless initial state with the help of an asymmetric 
potential.
In Sec. \ref{sec:prel} we describe the problem
and set up the definitions. In Sec.\ \ref{sec:floquet} we show how
the general solution of the problem at hand is constructed and
cover the cases of two simple resonances. In Sec.\
\ref{sec:halfinteger} a resonance of special interest is analyzed
by perturbative and numerical means. Finally, in 
Sec.\ \ref{sec:conclusions} we summarize and conclude.

\section{Preliminaries}
\label{sec:prel}
We wish to study the Schr{\"o}dinger equation 
\beq
i\kbar\frac{d\psi}{dt} = -\frac{\kbar^2}{2} \frac{d^2\psi}{dx^2} +
U(x) \psi \sum_{n=1}^{\infty} \delta(t-n).
\enq
This equation describes a particle subject to a periodic potential
$U(x)$ that is flashed on for short, periodically repeated pulses.
It can be realized with atoms in an optical lattice \cite{raizen,duffy,sadgrove};
the kicked rotor problem corresponds to the case $U(x)\propto \sin
x$, while the superposition of two sines has been predicted 
to result in a directed current \cite{lw1}. The parameter $\kbar$ is a 
dimensionless, effective Planck constant.
Units are chosen so that the spatial periodicity of the potential
is $2\pi$ and the temporal period of the flashing is unity.

The time development of the wave packet is given by
\beq\label{timeevolutionequation}
\psi(x,t+1) \equiv {\mathcal U}\left[\psi(x,t)\right] = e^{-iV(x)}
\int_{0}^{2\pi} dx' \sum_{k=-\infty}^{\infty} e^{-ik(x-x') -i\kbar
\frac{k^2}{2}} \psi(x',t),
\enq
where the operator $\mathcal U$ was implicitly defined. We assume
that the wave function has the same spatial periodicity as the
potential, and hence the momentum variable $k$ is restricted to
integers. The extension to non-integer momenta is straightforward 
but not important to the objectives of this paper. 
For notational convenience we have defined
$V(x)=U(x)/\kbar$.

The time development can be written in terms of the Floquet states
$w_j(x)$, which are eigenstates with associated quasi-energies
$\omega_j$ of the evolution operator for one temporal period:
\beq\label{floquetequation}
e^{-i\omega_j} w_j(x) = {\mathcal U}\left[w_j(x)\right].
\enq
Since $\mathcal U$ is unitary, the quasi-energies $\omega_j$ are
real. The time evolution of an initial state $\psi(x,0)$ is
constructed as follows:
\beq\label{timeevolutionresolved}
\psi(x,t) = \sum_j c_j
e^{-i\omega_j t} w_j(x),
\enq
with
\beq
c_j = \int dx w_j(x)^* \psi(x,0).
\enq
Note that since the evolution operator $\mathcal U$ propagates the
system for one unit of time $\Delta t=1$, Eq.\
(\ref{timeevolutionresolved}) is valid for integer $t$ only.

This type of analysis was first performed for the kicked rotor in 
Refs.\ \cite{izrailev,izrailevreview}. It was found that quantum 
resonances are associated with a banded quasienergy spectrum. 
In the present paper, the same type of analysis will be employed 
with a specific goal in mind:
It was found numerically in Ref.\ \cite{lw1} that when the
effective Planck constant $\kbar$ takes on values that are integer
or half-integer multiples of $\pi$, the ensuing resonant behavior
may result in a directed current. We shall now investigate how
this is reflected in the quasienergy spectrum.

\section{Floquet states at resonances}
\label{sec:floquet}
This section contains mainly a brief repetition and 
slight generalization of the findings of Ref.\ \cite{izrailev}, 
so that the formalism can be applied to the case of an 
asymmetric flashing potential. 
In order to get acquainted with the system and define the
terminology, we start with the simplest case, where the Planck
constant is an integer multiple of $4\pi$. It is well known
that in this case, the density stays unchanged at all
times, the mean momentum increase is zero and the energy increases
quadratically with time for any potential $V(x)$ 
\cite{duffy,daley,izrailev,izrailevreview}.

When $\kbar=4\pi$, then $\exp(-i\kbar k^2/2) = 1$ for all $k$, 
and the unitary operator $\mathcal U$ in Eq.\
(\ref{timeevolutionequation}) simplifies to
\beq
{\mathcal U}\left[\psi(x)\right] = e^{-iV(x)} \int_{0}^{2\pi}
dx' \sum_{k=-\infty}^{\infty} e^{-ik(x-x')} \psi(x',t) =
e^{-iV(x)} \psi(x,t).
\enq
Already from here we can see the solution to the full problem, but
in order to prepare ourselves for more complicated cases we 
solve for the time evolution using Floquet analysis. The eigenvalue
equation reads
\beq
e^{-iV(x)} w_j(x) = e^{-i\omega_j} w_j(x).
\enq
The phase factor on the left-hand side is space dependent, but
that on the right-hand side is not. This has a solution if
$w_j(x)$ is nonzero only for a discrete number of spatial points
$x$. The discrete index $j$ has to be changed into a continuous
one, $x_0$, and the solution for the continuous set of 
eigenstates, $\{w_{x_0}\}_{0\leq x_0<2\pi}$, 
is found to be
\beq
w_{x_0}(x) = \delta(x-x_0), \, \omega_{x_0} = V(x_0).
\enq

The expansion coefficients for the initial state are
\beq
c_{x_0} = \int dx w_{x_0}(x)^* \psi(x,0) = \psi(x_0,0),
\enq
and the wave function at integer time instances $t$ is
\beq
\psi(x,t) = \int dx_0 c_{x_0} e^{-i\omega_{x_0}}w_{x_0}(x) =
\psi(x,0) e^{-itV(x)}.
\enq
We conclude that the system remains unchanged at all times except
for a multiplicative phase factor; there is no transport.

We are now prepared to discuss the more general case
$\kbar=\pi/m$, where $m$ is an integer. (The case $\kbar=2\pi$ is more
easily solved by simpler means \cite{lw1}.) We have for the sum 
over momenta
\beq \label{deltasum_general}
\sum_{k=-\infty}^{\infty} e^{-i\pi
\frac{k^2}{2m}}e^{-ikx} = \sum_{j=0}^{2m-1} A_j
\delta(x-j\frac{\pi}{m}),
\enq
where
\beq
A_j = \frac{1}{2m} \sum_{n=0}^{2m-1}e^{-i\frac{\pi}{2m}(n^2-2nj)}.
\enq
The Floquet equation becomes
\beq \label{floqueteq_general}
e^{-i\omega_{x_0}}w_{x_0}(x) =
e^{-iV(x)} \sum_{j=0}^{2m-1} A_j w_{x_0}(x-j\frac{\pi}{m}).
\enq
By the same argument as above in the $\kbar=4\pi$ case, 
the solution must be of the form
\beq
w_{x_0}(x) = \sum_{l=0}^{2m-1} \alpha_{x_0 l}
\delta(x-x_0-l\frac{\pi}{m}).
\enq
We have come to the important conclusion that at a quantum 
resonance, each Floquet eigenstate is
nonzero only at a discrete set of points. As a result, the Floquet spectrum 
is divided into a number of discrete bands, within each of which the 
quasienergy depends on the continuous parameter $x_0$.
Now, inserting this ansatz
and equating the coefficients yields the equations for the
factors $\alpha_{x_0,l}$ 
\beq
e^{-i\omega_{x_0}} \alpha_{x_0,n}
= \sum_{l=0}^{2m-1} M_{nl}\alpha_{x_0,l},
\enq
where
\beq
M_{nl} = e^{-iV(x_0+n\pi/m)} A_{n-l}.
\enq
The indices $l$ and $n$ are modulo $2m$. There are $2m$ solutions
to this eigenvalue equation, which will be labeled 
$\alpha_{x_0l}^{\mu}$, with $\mu=0..2m$.
But there is a degeneracy, since we can choose phases such that
$w_{x_0+r\pi/m}^{\mu}(x) = w_{x_0}^{\mu}(x)$, or
$\alpha^{\mu}_{x_0+r\pi/m,l} = \alpha^{\mu}_{x_0,l+r}$. This
degeneracy needs to be taken care of by restricting $0\leq
x_0<\pi/m$, in order to avoid overcounting.

Let us now use these eigenstates to construct the long-time
development of the wavepacket. We specialize to the case of
homogeneous initial conditions, $\psi(x,0)=1/\sqrt{2\pi}$, whereby
the coefficients $c$ become
\beq
c^{\mu}_{x_0} = \frac1{\sqrt{2\pi}}\sum_{l=0}^{2m-1}
\alpha^{{\mu}*}_{x_0l},
\enq
and
\beq
\psi(x,t) = \frac1{\sqrt{2\pi}}\sum_{{\mu}}\sum_{l}
\alpha^{{\mu}*}_{x-l_x\pi/m,l} e^{-i\omega^{\mu}_{x-l_x\pi/m}t}
\alpha^{\mu}_{x-l_x\pi/m,l_x},
\enq
where $l_x$ is the integer that fulfills $\pi l_x/m \leq x <
\pi(l_x+1)/m$. After the completeness of the basis $w_{x_0}^{\mu}$ 
has been exploited, we are free to extend 
the definition of the amplitudes $\alpha$ 
to the whole range $0\leq x_0 < 2\pi$, and there results
\beq
\psi(x,t) = \frac1{\sqrt{2\pi}}\sum_{\mu}\sum_{l}
\alpha^{{\mu}*}_{x,l}\alpha^{\mu}_{x,0} e^{-i\omega^{\mu}_{x}t}.
\enq
The goal is to calculate the time dependence of the momentum. 
It is given by
\beq
\langle \psi(t) | \frac{dV}{dx} |\psi(t) \rangle = \frac1{2\pi}
\sum_{{\mu},{\mu}'} \sum_{ll'} \int_0^{2\pi} dx \frac{dV(x)}{dx}
e^{-i(\omega_x^{{\mu}}-\omega_x^{{\mu}'})t} \alpha^{{\mu}*}_{x,l}
\alpha^{{\mu}'}_{x,l'} \alpha^{{\mu}'*}_{x0} \alpha^{\mu}_{x0}.
\enq
The sum can divided into a part that oscillates in time and a
constant part. The constant part will be responsible for a linear
increase of the momentum, if there is one. This term is denoted 
by $F$ and is given by
\beq \label{force}
F = \frac1{2\pi}\sum_{{\mu}} \sum_{l,l'} \int_0^{2\pi} dx V'(x)
\alpha^{{\mu}*}_{x,l} \alpha^{{\mu}}_{x,l'}
|\alpha^{\mu}_{x,0}|^2.
\enq
This is the final expression for the rate of momentum increase for an
initially homogeneous wavepacket.

The solution method outlined above will now be applied to the
case $\kbar=\pi$, i.~e.\ $m=1$. We have $\exp(-i\pi
k^2/2) = 1$ for even $k$ and $-i$ for odd $k$, and the
kernel is
\beq
\sum_k e^{-ikx} e^{-i\pi k^2} = \frac{1-i}{2}\delta(x) +
\frac{1+i}{2}\delta(x-\pi).
\enq
The solutions to the Floquet equation consist of two bands, whose 
discrete indices $\mu$ will be labeled + and -. 
The eigenvectors are
\bea
\alpha^{\pm}_{x_0,0} = \left[ \frac{e^{-i\Delta
V}}{2}\left( 1 \mp \frac{\sin \Delta V}{\sqrt{1+\sin^2\Delta V}}
\right)\right]^{1/2},\nonumber\\
\alpha^{\pm}_{x_0,1} = \pm\left[ \frac{e^{i\Delta V}}{2}\left( 1
\pm \frac{\sin \Delta V}{\sqrt{1+\sin^2\Delta V}}
\right)\right]^{1/2},
\ena
with the eigenvalues
\beq
e^{-i\omega^{\pm}_{x_0}} =
\frac{1-i}{2}e^{-{\mathcal V}} \left[\cos \Delta V \pm i
\sqrt{1+\sin^2 \Delta V} \right],
\enq
where $\Delta V =
(V(x_0)-V(x_0+\pi)/2$, and ${\mathcal V} = (V(x_0)+V(x_0+\pi)/2$.
Thus,
\beq \omega_{x_0}^{\pm} = \frac{\pi}{4} +
{\mathcal V}(x_0) \mp \arctan \frac{\sqrt{1+\sin^2 \Delta
V(x_0)}}{\cos \Delta V(x_0)}. 
\enq

Now insert this into the general expression, Eq.\ (\ref{force}), 
for the time development.
There results for the time-independent part of the force
\bea
F &=& \sum_{+,-} \int_0^{2\pi} dx \frac{dV(x)}{dx}
\left|
\sqrt{\frac{e^{-i\Delta V}}{2} (1 \mp \frac{s}{\sqrt{1+s^2}})} \pm
\sqrt{\frac{e^{i\Delta V}}{2} (1 \pm \frac{s}{\sqrt{1+s^2}})}
\right|^2
\left|1 \mp \frac{s}{\sqrt{1+s^2}} \right|^2
\nonumber\\
&=& \int_{0}^{2\pi} dx V'(x)
\left(1 - \frac{sc}{1+s^2}\right),
\ena
where $s$ and $c$ are short for $\sin \Delta V(x)$ and $\cos \Delta V(x)$,
respectively. The presence of a term odd in $\Delta V(x)$ is crucial.
Now use the periodicity of $V$ to transform
\bea
F &=& \int_{0}^{2\pi} dx V'(x)
+ \frac12 \int_0^{2\pi} -V'(x)\frac{sc}{1+s^2}
+ V'(x+\pi)\frac{sc}{1+s^2}
\nonumber\\
&=& \int_{0}^{2\pi} dx V'(x)
- \int_0^{2\pi} dx \Delta V'(x){dx}\frac{sc}{1+s^2}
\nonumber\\
&=& \int_{x=0}^{2\pi} dV(x) - \int_{x=0}^{2\pi} d(\Delta V)
\frac{\sin \Delta V\cos\Delta V}{1+\sin^2\Delta V} = 0,
\ena
by
the periodicity of the potential. This could be
done because the integrand of the second term depends on the
coordinate solely through the potential difference $\Delta V$.
This concludes the demonstration that the drift is zero in an
initially homogeneous system for any potential $V$ at the
resonance $\kbar=\pi$.

\section{Resonance at $\kbar=\pi/2$}
\label{sec:halfinteger}

The half-integer resonances are especially interesting since they are known
to result in directed transport even if the initial state is homogeneous
\cite{lw1}. We therefore study the case $\kbar=\pi/2$ with special care.
The momentum sum is now
\beq
\sum_{k}e^{-ikx}e^{-i\pi k^2/4} =
\frac12\left[e^{-i\pi/4} \delta(x) + \delta(x-\pi/2)
-e^{-i\pi/4} \delta(x-\pi) + \delta(x-3\pi/2)
\right],
\enq
and correspondingly the matrix $M$ for the Floquet eigenvectors reads
\beq
M = V \frac12\left(\begin{array}{cccc}
e^{-i\pi/4} & 1 & -e^{-i\pi/4} & 1 \\
1 & e^{-i\pi/4} & 1 & -e^{-i\pi/4} \\
-e^{-i\pi/4} & 1 & e^{-i\pi/4} & 1 \\
1 & -e^{-i\pi/4} & 1 & e^{-i\pi/4}\end{array}
\right),
\enq
where
\beq
V = {\rm diag}\left(e^{-iV(x)}, e^{-iV(x-\pi/2)}, e^{-iV(x-\pi)},
e^{-iV(x-3\pi/2)}\right).
\enq
We first solve this system by perturbative means. The quasienergies 
and eigenvectors to zeroth order in the potential $V(x)$ are
\bea
\omega^0_{(0)} = \pi, & \vec{\alpha}_{(0)}^0 = (-1,1,-1,1)^T,
\nonumber\\
\omega^1_{(0)} = 0, & \vec{\alpha}_{(0)}^1 = (1,1,1,1)^T,
\nonumber\\
\omega^2_{(0)} = -\pi/4, & \vec{\alpha}_{(0)}^2 = (0,-1,0,1)^T,
\nonumber\\
\omega^3_{(0)} = -\pi/4, & \vec{\alpha}_{(0)}^3 = (-1,0,1,0)^T,
\ena
so the correction to first order is
\bea
\vec{\alpha}_{(1)}^0 &=& \frac12i(-V_0+V_1-V_2+V_3)(1,1,1,1)^T +
\frac{e^{-i3\pi/8}}{\sqrt{2+\sqrt{2}}}
(\Delta_0,-\Delta_1,-\Delta_0,\Delta_1)^T,
\nonumber\\
\vec{\alpha}_{(1)}^1 &=& \frac12i(-V_0+V_1-V_2+V_3)(1,-1,1,-1)^T +
\frac{e^{i\pi/8}}{\sqrt{2-\sqrt{2}}}
(-\Delta_0,-\Delta_1,\Delta_0,\Delta_1)^T,
\nonumber\\
\vec{\alpha}_{(1)}^2 &=& -\sqrt{2}\Delta_1(e^{i\pi/4},1,e^{i\pi/4},1)^T,
\nonumber\\
\vec{\alpha}_{(1)}^3 &=& -\sqrt{2}\Delta_0(1,e^{i\pi/4},1,e^{i\pi/4})^T,
\ena
where we defined $V_j=V(x+j\pi/2)$ for $j=0,1,2,3$;
$\Delta_0=V_0-V_2$; and $\Delta_1=V_1-V_3$.
These eigenvectors can now be inserted into the expression for the force,
Eq.\ (\ref{force}). In order to obtain a closed expression, one has to 
make an assumption about the potential. We make the physically motivated 
choice
\beq
\label{sinepotential}
V(x) = U(\sin x + a \sin 2x),
\enq 
which models an atom subject to standing laser waves \cite{lw1}.
There results for small $U$
\beq
F = (3 + 2\sqrt2) a U^3 -
 (1 + \sqrt2)a^3 U^5.
\enq

\begin{figure}
\includegraphics[width=0.49\textwidth]{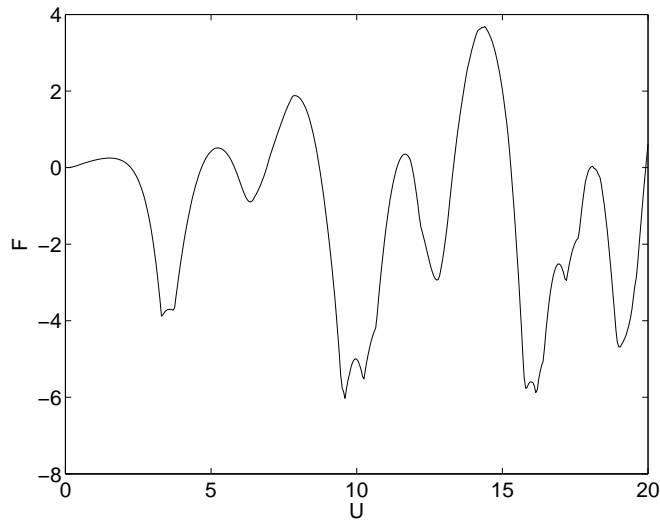}
\caption[]{rate of momentum increase on an initially homogeneous 
wavefunction for $a=0.3$, as a function of potential strength 
$U$, assuming the 
form of Eq.\ (\ref{sinepotential}) for the potential.
\label{fig:force}}
\end{figure}

\begin{figure}
\includegraphics[width=0.49\textwidth]{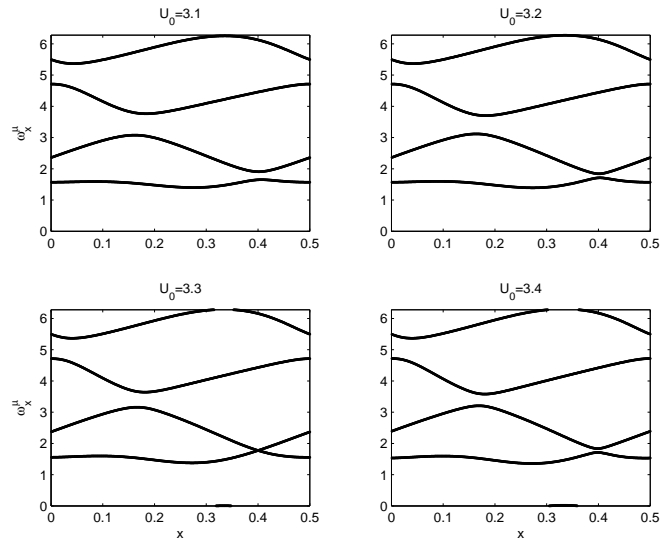}
\caption[]{Floquet eigenvalues $\omega_{x_0}^{\mu}$ as functions of $x_0$ 
for four choices of potential strength $U$. There appears a level crossing 
at $x\approx 0.4$ for $U=3.3$ which disappears again for larger $U$; 
this signals a sharp feature in the curve for the force $F$ in 
Fig. \ref{fig:force}. 
Because of periodicity, only the range $0<x<\pi/2$ is displayed.
\label{fig:freqs}}
\end{figure}

In order to go beyond perturbation theory, the $\kbar=\pi/2$ 
problem has to be solved numerically. Again we assume
the form Eq.\ (\ref{sinepotential}) for the potential.
Figure \ref{fig:force} depicts the numerically calculated force $F$ 
as a function of potential strength $U$, for the choice $a=0.3$. 
The force rises initially as the third power of $U$ as the 
perturbative calculation indicated, but for longer times the time 
dependence displays an approximate periodicity of $F$ 
with period $\pi$ as a function of $U$. 
We shall now see that it can be related to 
the structure of the Floquet spectrum. 
In Fig.\ \ref{fig:freqs}, we
display the numerically obtained eigenvalues $\omega_{x_0}^{\mu}$ 
for a range of values of $U$. 
It is seen that a level crossing as a function of the continuous 
parameter $x_0$ appears when $U=3.3$ and then separates when $U$ is 
increased. At this point, two energy bands momentarily merge into one. 
This crossing is reflected in Fig.\ \ref{fig:force} as a 
pointed feature in the force curve $F(U)$. Closer inspection reveals 
that there is a discontinuity in the first derivative at the 
turning point, but there does not appear to be a cusp.

Level crossings in Floquet spectra have been seen in various contexts 
to be associated with resonance phenomena \cite{levelcross1, levelcross2}. 
However, in Refs.\ \cite{levelcross1,levelcross2}, the situation is 
different: the eigenvalue spectra are discrete and the level crossings 
occur as a control parameter is varied. In the 
present system, the eigenvalues form a continuous set and the level 
crossings take place in the two-dimensional space formed by the 
continuous index $x_0$ and the parameter $U$; in other words, the 
crossing is a merging of two quasienergy bands. 
The crossings are not 
easily interpreted as a resonance phenomenon, but they are connected 
with the sharp turning points of the force $F$ as a function of $U$
that can be seen in Fig.\ \ref{fig:force}. In fact, every such 
sharp turning point coincides with a level crossing, i.~e., a 
merging of bands, in the continuous spectrum.
In Fig.\ 3, we see how a number of level crossings appear around $U=19$. 
\begin{figure}
\includegraphics[width=0.49\textwidth]{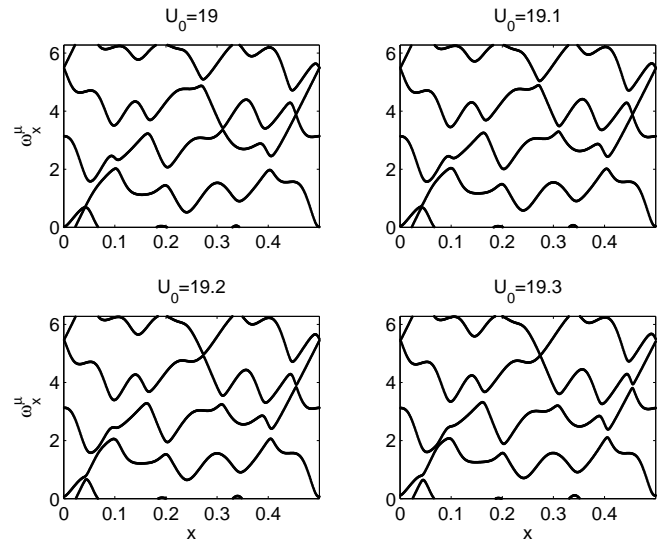}
\caption[]{Floquet eigenvalues $\omega_{x_0}^{\mu}$ as functions of $x_0$ 
for four choices of potential strength $U$. 
\label{fig:freqs2}}
\end{figure}
Here, the minimum in the curve $F(U)$ is smoother, and 
accordingly it 
is associated with a {\it sequence} of level crossings instead of just one.

\begin{figure}
\includegraphics[width=0.49\textwidth]{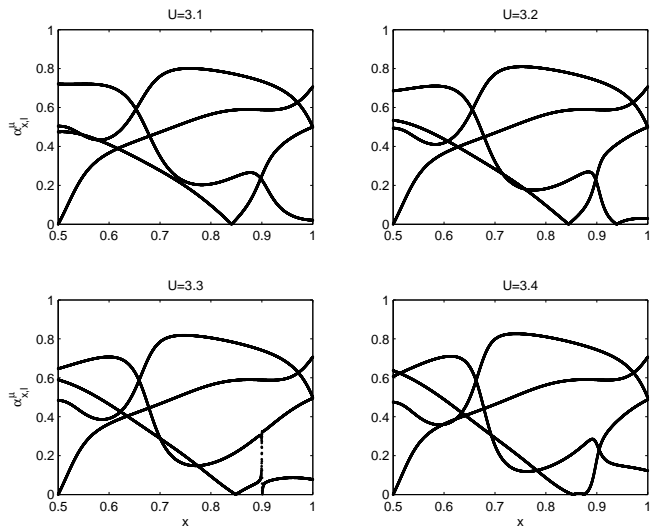}
\caption[]{A selection of eigenvector components $\alpha_{xl}^{\mu}$ 
as functions of $x$, for a few choices of $U$.
\label{fig:evecs}}
\end{figure}

A straight level crossing originates from two eigenvectors 
$\vec{\alpha}^{\mu}$ that have different symmetry in the sense that the 
overlap with respect to a small change $\delta U$ of the control 
parameter $U$ vanishes;
\beq
(\vec{\alpha}^{\mu_1})^T \frac{dA}{dU} \delta U \vec{\alpha}^{\mu_2} = 0.
\enq
(Alternatively, a small variation in the index $x_0$ can be 
considered, since the level crossing takes place in the two-dimensional 
space of $U, x_0$.) The generic situation is that the overlap is 
nonzero, which results in an avoided crossing and a mixing of the 
eigenvectors; when the overlap vanishes the energies cross and the 
eigenvectors do not mix.
Indeed, this decoupling of eigenvectors is clearly seen in the plots of 
the components 
$\alpha$ of the eigenvectors, 
a selection of which is displayed in Fig.\ \ref{fig:evecs}.
The avoided level crossing is accompanied by the crossing 
at $x\approx 0.9$ of the components of the corresponding eigenvectors. 
Close to  
the turning point the slope gets steeper and eventually the two lines 
do not cross at all, indicating that the overlap 
between the modes vanishes. This is, in turn, 
reflected in the force integral (\ref{force}) and manifests itself 
in the sharp bending of the curve.

\section{Conclusions}
\label{sec:conclusions}
We have investigated the Floquet spectrum for a wavepacket subject to 
periodic forcing, with special attention to quantum resonances, 
where the effective Planck constant is equal to 
a rational multiple of $\pi$ and the quasienergies form a band structure. 
We derived
an expression for the force on the wavepacket in terms of Floquet 
eigenstates and concluded that a nonzero mean velocity is obtained 
from homogeneous initial conditions only at minor resonances when the 
Planck constant is equal to a half-integer multiple of $\pi$. 
The oscillatory dependence of the current on the potential strength 
was investigated for the special case of a potential composed of two 
harmonics, and it was seen how turning points in the curve of 
drift as a function of potential strength arise from level crossings in the 
Floquet spectrum.

\section{acknowledgments}
This work was supported by the Swedish Research Council.
The author is grateful to Mats Wallin for helpful discussions.


\begin{thebibliography}{99}
\bibitem{raizen} C.~F.\ Bharucha, J.\ C.\ Robinson, 
  F.\ L.\ Moore, Bala Sundaram, Qian Niu, and M.\ G.\ Raizen, 
  Phys.\ Rev.\ E {\bf 60}, 3881 (1999).
\bibitem{duffy}  G.\ Duffy, S.\ Parkins, 
  T.\ M{\"u}ller, M.\ Sadgrove, R.\ Leonhardt, 
  and A.\ C.\ Wilson, Phys.\ Rev.\ E {\bf 70} 056206 (2004);
  G.~J.\ Duffy, A.~S.\ Mellish, K.~J.\ Challis, and A.~C.\ Wilson,
  Phys.\ Rev.\ A {\bf 70}, 041602(R) (2004).
\bibitem{sadgrove} M.\ Sadgrove, S.\ Wimberger, S.\ Parkins, 
  and R.\ Leonhardt, Phys.\ Rev.\ Lett.\ {\bf 94}, 174103 (2005).
\bibitem{fishman} S.\ Fishman, D.~R.\ Grempel, and R.~E.\ Prange, 
  Phys.\ Rev.\ Lett.\ {\bf 49}, 509 (1982).
\bibitem{izrailev} F.~M.\ Izrailev and D.~L.\
  Shepelyanskii, Theor.\ Math.\ Phys.\ {\bf 43},
  553 (1980).
\bibitem{izrailevreview} F.~M.\ Izrailev, Phys.\ Rep.\ 
  {\bf 196}, 299 (1990).
\bibitem{daley} A.~J.\ Daley and A.~S.\ Parkins,
  Phys.\ Rev.\ E {\bf 66}, 056210 (2002).
\bibitem{wimberger} S.\ Wimberger, I.\ Guarneri, and 
  S.\ Fishman, Phys.\ Rev.\ Lett.\ {\bf 92}, 084102 (2004); 
  Nonlinearity {\bf 16}, 1381 (2003).
\bibitem{dd} I.\ Dana and D.~L.\ Dorofeev, Phys.\ Rev.\ E, 
  to be published (nlin.CD/0509035).
\bibitem{lw1} Emil Lundh and Mats Wallin, Phys.\ Rev.\ Lett.\ {\bf 94}, 
  110603 (2005). 
\bibitem{levelcross1} M.\ Latka, P.\ Grigolini, and B.~J.\ West, 
  Phys.\ Rev.\ A {\bf 50}, 1071 (1994).
\bibitem{levelcross2} T.\ Timberlake and L.~E.\ Reichl, 
  Phys.\ Rev.\ A {\bf 59}, 2886 (1999).
\end{thebibliography}
\end{document}